\documentclass[aps,pra,superscriptaddress,twocolumn]{revtex4}
\usepackage{amsmath,epsfig,mathrsfs,graphicx,amssymb,color}
\usepackage[T1]{fontenc}
\usepackage{t1enc}
\usepackage{hyperref}
\usepackage{array}
\usepackage{booktabs}

\def\e{\boldsymbol e}

\newcommand{\g}[1]{{\boldsymbol #1}}

\def\be#1\ee{\begin{equation}#1\end{equation}}
\newcommand{\ba}{\begin{eqnarray} }
\newcommand{\ea}{\end{eqnarray} }

\begin{document}

\title{Optimal classical and quantum real and complex dimension witness}
\author{Josep Batle}
\affiliation{Institut IES Can Peu Blanc, C. Ronda Nord 19, 07420 sa Pobla, Balearic Islands, Spain}
\affiliation{Departament de F\'{\i}sica, Universitat de les Illes Balears, 
                07122 Palma de Mallorca, Balearic Islands, Spain}
\email{jbv276@uib.es, batlequantum@gmail.com}
\author{Adam Bednorz}
\affiliation{Faculty of Physics, University of Warsaw, ul. Pasteura 5, PL02-093 Warsaw, Poland}
\email{Adam.Bednorz@fuw.edu.pl}


\begin{abstract}
We find the minimal number of independent preparations and measurements certifying the dimension
of a classical or quantum system limited to $d$ states, optionally reduced to the real subspace. As a dimension certificate, we use the linear independence
tested by a determinant. We find the sets of preparations and measurements that maximize the chance to detect larger space if the extra contribution is very small.  
We discuss the practical application of the test to certify the space logical operations on a quantum computer.

\end{abstract}

\maketitle

\section{Introduction}
Few-state systems have become standard building blocks in current classical and quantum technologies. In particular
two-state (bit) classical logic is the base of computers and information transfer. Qubits (quantum two-state systems)
are basic logical elements of quantum computers. To increase the quality of classical and quantum computation and communication,
these systems need precise certification. For instance, the contribution of external states to qubit operations can lead to systematic errors,
accumulated in long operation circuits and difficult to correct.

To certify that the number of classical or quantum states is limited, one can use a dimension witness (the dimension is the number of states). 
The usual construction of the witness is based on the two-stage protocol, the initial preparation and final measurement \cite{gallego}, which are taken from several respective possibilities, and are independent of each other. Importantly, the preparation phase must be completed before the start of the measurement. Such early witnesses were based
on linear inequalities, tested experimentally \cite{hendr,ahr,ahr2,dim1} but they could not detect e.g. small contributions from other states. In the latter case, it would be better to use a nonlinear witness \cite{leak}.
 A completely robust witness must be based on equality, i.e.,
a quantity, which is exactly zero up to a certain dimension and can be nonzero above \cite{dim,chen}. 

A good witness test is the linear independence of the  specific outcome probability $p(y|x)$ for the preparation $x$ and measurement $y$ by a suitable determinant \cite{dim, chen}. 
In previous works, a witness of dimension $d$ needed $2k$ preparations and $k$ measurements, with $d\leq k$ for the classical system and $d^2\leq k$ for the quantum system. 
Equality-based tests, like the Sorkin equality \cite{sorkin} in the three-slit experiment \cite{tslit,btest1,btest2} 
 testing Born's rule \cite{born}, 
belong to a family of precision tests of quantum mechanics, benchmarking our trust in fundamental quantum models and their actual realizations.

Here we show that the number of preparations can be reduced to $k+1$, preserving the properties of the witness, i.e., being zero for $k\geq d,
(d+1)d/2, d^2$ for classical, real, and complex quantum systems, respectively.  The real quantum system is described by a Hilbert subspace of only real vectors, 
which occur, e.g., when the Hamiltonian is purely imaginary and the unitary operations become real rotations in real space \cite{real}. 
The witness is essentially a determinant of the matrix with entries $p(y|x)$ and ones in the last row.
The construction of the witness allowed us to analyze both its extremal violation by additional states and minimal deviations. The extremal examples for
a bit, trit, or qubit can be described in the hybrid analytical-numerical form, while only the numerical form can be used for qutrits and higher dimensionw.
This analysis should help us to estimate the bounds in practical tests of the dimension and in the effort to eliminate parasitic states.

\begin{figure}
\includegraphics[scale=.3]{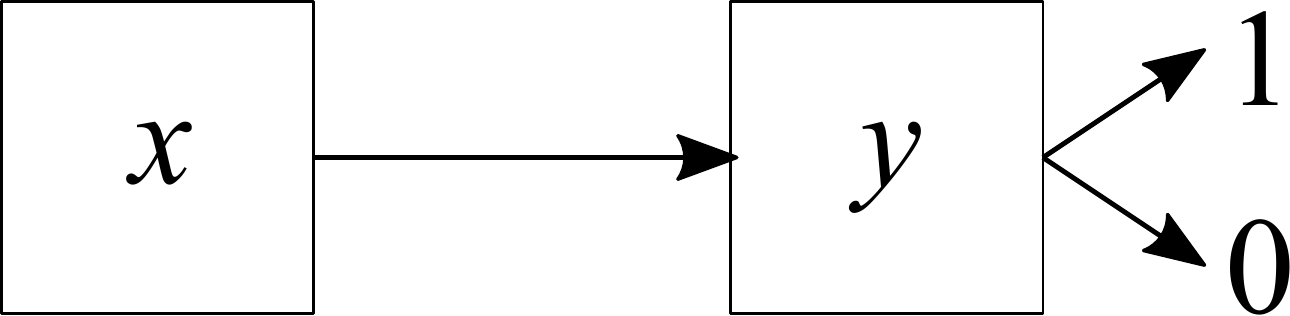}
\caption{Preparation and measurement scenario; the state is prepared as $x$ and measured by $y$ to give an outcome of either $1$ or $0$.}
\label{pms}
\end{figure}

\section{ Dimension certificate}
We consider the standard prepare and measure scenario with binary (yes or no or 1 or 0) outcome (Fig. \ref{pms}). 
The state is prepared in one of $m$ possibilities represented by Hermitian matrices 
$0\leq \hat{X}_1,...,\hat{X}_{m}$, $\mathrm{Tr}\hat{X}_j=1$.
The probability of the measurement of an outcome of $1$ (yes) is applied out of $k$ possibilities  
$p(y_i|x_j)\equiv p_{ij}=\mathrm{Tr}\hat{Y}_i\hat{X}_k$ for the measurement operators $0\leq \hat{Y}_1,...,\hat{Y}_k\leq \hat{1}$.
If the system is classical, with $d$ states,
then $p_{ij}=\sum_{a}q_{ia}r_{aj}$  with $a=1,...,d$ and $r$, $q$ describing the transfer probabilities from the prepared state to the classical $d$-dimensional register while $q$ is the transfer probability from the register to the measurement outcome.
The quantum states and measurements can be either real or fully complex.
The real register of a $d-$level quantum state consists of $(d+1)d/2$ Gell-Mann basis matrices with all zeros except a single $1$ on 
the diagonal or a symmetric off-diagonal pair of two $1$'s and the complex register is enlarged by $(d-1)d/2$ antisymmetric matrices with entries $(i,-i)$.

{\bf Main theorem.} Suppose a witness of dimension $d$,  $W(\{p_{ij}\})$, is always equal to zero for a system of dimension $\leq d$. Then
the minimal number of measurements is $k=d^2$ and $k=(d+1)d/2$ in the complex and real spaces, respectively, while the minimal number of prepared states is $m=k+1$.

{\it Proof.} We set $k=d^2$ and $k=(d+1)/2$ in the complex and real spaces, respectively. Then each operator $\hat{Y}_i$ has $k$ 
independent entries while $\hat{X}_j$ has $k-1$ independent entries (the trace is 1). If either the number of measurements is $k'<k$ or the number of preparations is $m<k+1$, then 
the number of all $p_{ij}$ is $k'm$, and the dimension of available space is not smaller, either $k'k$ or $m(k-1)$, meaning that $p_{ij}$ are essentially independent.

On the other hand, setting $m=k+1$,  and $\hat{Y}_{k+1}=\hat{1}$ (the auxiliary always-yes measurement, which is actually not performed, meaning $p_{k+1,j}=1$) to get a $(k+1)\times(k+1)$ matrix of $p_{ij}$ we can define the witness
\be
W_k=\det p\label{dett}
\ee
Note that the determinant can be reduced to a $k\times k$ matrix, without changing its value, subtracting e.g. the last column from each previous one, i.e.
\be
p_{ij}\to p_{ij}-p_{i,k+1}
\ee
to make it equivalent to the previous one, with all odd preparations being identical \cite{dim}. In our way, the number of preparations is reduced from $2k$ to $k+1$. Nevertheless,  the original $(k+1)\times (k+1)$ form is better suited for the analysis of the extremal cases. 
Now, $W_k=0$ for $d\leq k$ classically, $(d+1)d/2\leq k$ in the real quantum case and $d^2\leq k$ in the complex quantum case.
The determinant is a sum of signed products of permuted elements of every column. 
This means that the given column or row can occur only once in a single product.
Therefore, whenever the prepared state $\hat{X}_i$ is a convex combination of pure states, the determinant is also a convex combination
of the cases with $\hat{X}_i$ replaced by the pure states. Looking for the extreme case, one can reduce the search to pure states.
For the measurement $\hat{Y}_j$ the situation is a bit different, the search can be reduced to projections 
but not necessarily to one dimension.
Because of the always-yes measurement, it suffices to consider  the maximal dimension of the projection space up to $d/2$.

\begin{table}
\begin{tabular}{*{10}{c}}
\toprule
$k$&1&2&3&4&5&6&7&8&9\\
\midrule
$\mathrm{max}W_k$&1&1&2&3&5&9&32&56&144\\
\bottomrule
\end{tabular}
\caption{The classical maximum of $W_k$ equivalent to a maximal determinant of the $k\times k$ matrix with entires of $0$ or $1$.}
\label{tabone}
\end{table}

\begin{table*}
$$
\begin{pmatrix}
1&0\end{pmatrix},\:
\begin{pmatrix}
1&0&0\\
0&1&0\end{pmatrix},\:
\begin{pmatrix}
1&0&0&1\\
0&1&0&1\\
0&0&1&1\end{pmatrix},\:
\begin{pmatrix}
1&0&0&0&1\\
0&1&0&0&1\\
0&0&1&0&1\\
0&0&0&1&1
\end{pmatrix},\:
\begin{pmatrix}
1&0&0&1&1&0\\
0&1&0&1&1&0\\
0&0&1&1&1&0\\
0&0&0&1&0&1\\
0&0&0&0&1&1\end{pmatrix},$$
$$
\begin{pmatrix}
1&1&0&1&0&0&0\\
0&1&1&0&1&0&0\\
0&0&1&1&0&1&0\\
1&0&0&1&1&1&0\\
0&1&0&0&1&1&0\\
1&0&1&0&0&1&0\end{pmatrix},\:
\begin{pmatrix}
1&0&1&0&1&0&1&0\\
1&1&0&0&1&1&0&0\\
1&0&0&1&1&0&0&1\\
1&1&1&1&0&0&0&0\\
1&0&1&0&0&1&0&1\\
1&1&0&0&0&0&1&1\\
1&0&0&1&0&1&1&0
\end{pmatrix},
$$
$$
\begin{pmatrix}
1&0&1&0&0&1&1&0&0\\
1&1&0&1&0&0&1&1&0\\
1&1&1&0&1&0&0&1&0\\
0&1&1&1&0&1&0&0&0\\
0&0&1&1&1&0&1&0&0\\
1&0&0&1&1&1&0&1&0\\
0&1&0&0&1&1&1&0&0\\
0&0&1&0&0&1&1&1&0
\end{pmatrix},\;
\begin{pmatrix}
1&1&0&0&0&0&0&0&1&1\\
1&0&1&0&0&0&0&1&0&1\\
1&0&0&1&0&0&1&0&0&1\\
1&0&0&0&1&1&0&0&0&1\\
1&0&0&0&1&0&1&1&1&0\\
1&0&0&1&0&1&0&1&1&0\\
1&0&1&0&0&1&1&0&1&0\\ 
1&1&0&0&0&1&1&1&0&0\\
1&1&1&1&1&0&0&0&0&0
\end{pmatrix}
$$
\caption{Binary matrices $(k+1)\times (k+1)$ (without the last row of $1$'s) with the maximal $W_k$
given in Table \ref{tabone}. It reduces to a $k\times k$ determinant for example  the last (not shown) row
is subtracted 
from those rows with $1$ in the first column, reversing the sign of those rows. }\label{tabtwo}
\end{table*}

\begin{table*}
\begin{tabular}{*{12}{c}}
\toprule
$k$&2r&2c&3r&3c&4r&4c&5r&5c&6r&6c\\
\midrule
1&1&&&&&&&&&\\
2&0.65&&1&&&&&&&\\
3&0&0.38&0.84&&2&&&&&\\
4&0&0&0.60&0.63&1.87&&3&&&\\
5&0&0&0.42&0.46&1.78&&3.14&&6&\\
6&0&0&0&0.33&1.61&1.68&3.40&&5.04&\\
7&0&0&0&0.23&1.41&1.64&3.51&3.72&6.05&6.18\\
8&0&0&0&0.15&1.30&1.47&3.65&3.79&7.49&7.50\\
9&0&0&0&0&1.29&1.39&3.77&3.84&10.14&10.34\\
\bottomrule
\end{tabular}
\caption{The quantum maximum of $W_k$ for the $(k+1)\times (k+1)$ matrix for a $d$-dimensional system, either real (r) or complex (c).
An empty cell means the value is the nearest number to the left in the row.}
\label{taboneq}
\end{table*}

\section{ Maximal nonzero value}
The relevant question about the dimension witness is how non-zero it is, which allows determining whether the system has the desired dimension.
First of all, there always exists a classical maximum, by linearity obtained for $p_{ij}$ equal to $0$ or $1$, known as the Hadamard determinant. The general upper bound is $(k+1)^{(k+1)/2}/2^k$,
but is not reached for particularly low values of $k$, for which the maximum can be found using algebraic methods \cite{sloane,ehr,woj}, as summarized in Table \ref{tabone}.
Second, for a classical system, the maximum is achieved immediately when $d>k$, taking the initial classical state  $m=1$, $r_{m1}=1$ and $0$ otherwise, and $q_{im}=p_{ij}$ for a \emph{given} $p$ 
(a maximal example is given in Table \ref{tabtwo}).
Third, the above classical maximum is also obtained for a quantum state with $d>k$, using analogous reasoning.
The nontrivial bounds are for a quantum system such that $d^2>k\geq d$. Due to he antisymmetry of the determinant, the bounds are always symmetric.

A special case is $k+1=(d+1)d/2$ and $k+1=d^2$ for  real and complex quantum states of dimension $d$, respectively.
In these cases the determinant can be written as a product of determinants of separate square matrices for the preparations $\hat{X}_i$ and measurements
$\hat{Y}_j$ so that (a) the maximum can be determined separately for $x$ and $y$ and (b) the same maximum is reached if either preparations or measurements are rotated by an arbitrary orthogonal/unitary matrix in the real/complex case, i.e $\hat{X}_i\to \hat{R}\hat{X}_i\hat{R}^T$ or
$\hat{X}_i\to\hat{U}\hat{X}_i\hat{U}^\dag$.

The maximum $W_2=(3/4)^{3/2}\simeq 0.65$ in both real and complex qubit space is reached by
\ba
&&\g{x}_1=(0,0,1),\nonumber\\
&&\g{x}_2=(\sqrt{3}/2,0,-1/2),\label{tri}\\
&&\g{x}_3=(-\sqrt{3}/2,0,-1/2),\nonumber
\ea (i.e., the vertices of the equilateral triangle) and $\g{y}_1=(0,0,1)$, $\g{y}=(1,0,0)$ using the Bloch sphere notation
\be
2\hat{Y}=\hat{1}+\g{y}\cdot\hat{\g{\sigma}},\:2\hat{X}=\hat{1}+\g{x}\cdot\hat{\g{\sigma}}\label{sig}
\ee
with $|\g{y}|=|\g{x}|=1$ and standard Pauli matrices $\hat{\g{\sigma}}=(\hat{\sigma}_1,\hat{\sigma}_2,\hat{\sigma}_3)$, where
$\hat{\sigma}_1=|1\rangle\langle 2|+|2\rangle\langle 1|$, $\hat{\sigma}_2=i|2\rangle\langle 1|-i|1\rangle\langle 2|$, and 
$\hat{\sigma}_3=|1\rangle\langle 1|-|2\rangle\langle 2|$; and $\hat{1}=|1\rangle\langle 1|+|2\rangle\langle 2|$.

For a complex qubit, the maximum $W_3=2\sqrt{3}/9\simeq 0.38$ is achieved for
\ba
&&\g{y}_1=(1,0,0),\:\g{y}_2=(0,1,0),\:\g{y}_3=(0,0,1)\nonumber\\
&&\g{x}_1=(1,1,1)/\sqrt{3}\nonumber\\
&&\g{x}_2=(1,-1,-1)/\sqrt{3}\\
&&\g{x}_3=(-1,-1,1)/\sqrt{3}\nonumber\\
&&\g{x}_4=(-1,1,-1)/\sqrt{3}\nonumber
\ea
 i.e. axes for $y$ and the vertices of the regular tetrahedron for $x$.

For higher $k$ and $d$, the search for the maximum of the determinant-based dimension witness becomes an arduous task if tackled only from the analytic point of view, 
although it is always some algebraic number. That is why we shall resort 
to numerical computations in order to maximize the determinant of the corresponding $(k+1)\times(k+1)$ matrix.  
Therefore, approximated or heuristic methods are required in practice to find the optimal states. The most successful statistical method to 
date -- the one we employ, is the stochastic model of {\it simulated annealing}  \cite{kirkpatrick83}, that is, the
Metropolis Monte Carlo algorithm with a fixed temperature $T$ at each state of the annealing schedule. 
Other methods exist which are not of a statistical nature, such as the downhill or amoeba or gradient methods \cite{Avriel}, which involve finite differences when considering 
the corresponding function in terms of all real variables involved.

The optimization is taken over the states $x_1,...,x_{k+1}$ and measurements $y_1,...,y_k$, which are real or complex unit vectors in $\mathbb{R}^d$, and $\mathbb{C}^d$, 
depending on the particular instance 
(e.g. $d=2$ refers to qubits, $d=3$ corresponds to qutrits and so on).  For $d=4,5$ we add a second set of vectors $y'_1,...,y'_k$   (and a third one, $y''$, for $d=6$) and Gram-Schmidt orthogonalize
 it to $y$, covering all possible dimensions of the projection space in a single run. With a proper parametrization, the problem consists of finding the supremum of the determinant (\ref{dett}). Initially, the temperature is set to the high value $T_0$, which implies that the domain of possible values for the variables $\Omega$ is broadly spread. 
Finishing one cycle means visiting all the variables in $\Omega$ one after the other. After that, we compute $W_k$, which is the cost function. Then, the cycle starts  anew with a 
different temperature $T_1$ (we choose the temperature to decrease as $T_{s+1} = r\, T_s$, $r=1/4$ with $s$ being the number of runs). As the temperature drops, the 
domain $\Omega_s$ continuously decreases until we reach the desired precision ($10^{-9}$ in our case), that is, the algorithm terminates when some stopping criterion is met. 
The details of the algorithms, their use and he results are given in Appendix A. 

The numerical results are presented in Table \ref{taboneq}. 
We found a few remarkable features: (i) The maximum needs the maximal dimension of the projection space, 
i.e., $2$ for $d=4,5$ and $3$ for $d=6$. (ii) For $d$ close to $k$, the real and complex case give the same maximum so that the witness cannot distinguish them. 
(iii) For some maxima the corresponding states and measurements are surprisingly regular (e.g., heptagonal symmetry for $k=7$, $d=5$ in the complex case), and the maximum is rational, 
while others are almost completely irregular (e.g. $k=5$, $d=3$ in the complex case), (iv)
The value increases with $d$ but not always with $k$.

\section{Detection of higher dimension}
The most practical application of the determinant-based witness is the diagnostic test of a finite-dimensional quantum system.
Suppose the system is designed to be a perfect $d$-level state and we want to check it with high accuracy.
A possible contribution from an imaginary  part (for a real state) or a higher level (a complex state) is expected to be small, 
so even a small nonzero value of $W_k$ would detect this.
Making a decomposition
\be
\hat{Y}_i=\hat{Y}_i^{0}+\delta\hat{Y}_i,\:
\hat{X}_j=\hat{X}_j^{0}+\delta\hat{X}_j
\ee
with $\hat{Y}^{0}$ and $\hat{X}^{0}$ restricted to a $d$-dimensional Hilbert space (real or complex) and only (small) deviations $\delta\hat{Y}$ and $
\delta\hat{X}$ in the imaginary or higher dimensional states.  To detect higher (imaginary) states one must use the witness for $k\geq (d^2+d)/2$ (real case)  or $k\geq d^2$ (complex case) 
because lower $k$ will give a nonzero value even for clean $d$ levels. The Jacobi identity implies in the lowest perturbative order
\be
W_k\simeq \mathrm{Tr}\; \delta p\mathrm{Adj}\: p^{0}\label{dev}
\ee
where $\mathrm{Adj}\: p$ is the adjoint matrix of $p_{ij}^{0}=\mathrm{Tr}\hat{Y}_i^{0}\hat{X}_j^{0}$
and $\delta p_{ij}=\mathrm{Tr}\delta\hat{Y}_i\delta\hat{X}_j$.
For $k>(d+1)d/2$ or $k>d^2$ the  all the minors in the zeroth order  are zero so the optimal choice is $k=(d+1)d/2,d^2$.
Moreover, as stressed earlier for this particular $k$, the values of the minors do not depend on the basis of $x$ or $y$.
To maximize the witness with respect to a potential higher-space contribution one has to maximize the minors. In particular, if
a single preparation and measurement are suspected, we simply maximize the corresponding minor, taking the determinant with
$k\to k-1$ whose maximum we have already discussed.

The practical application of the test requires $N$ repetitions of $k^2+k$ experiments for all values of $i$ and $j$,
with $p_{ij}\to N_{ij}/N$ for $N_{ij}$ positive results.
Assuming independence between experiments, the variance of $W_k$ in the null case (a perfect $d$-level state) can then be estimated as
\be
N\langle W_k^2\rangle\simeq \sum_{ij}p_{ij}(1-p_{ij})(\mathrm{Adj}\: p)_{ji}^2
\ee
For large $N$ this equation allows us to estimate how perfect the qubit (or any $d$-dimensional system) is.
For $d=2$ (complex), we numerically found the constraint $\langle W_4^2\rangle\leq 1/6N$, saturated by preparations and measurements given by
\ba
&&\g{x}_1=\g{x}_2=(0,0,-1),\: \g{x}_3=(2\sqrt{2}/3,0,1/3)\nonumber\\
&&\g{x}_{4,5}=(-\sqrt{2}/3,\pm\sqrt{2/3},1/3)\\
&&\g{y}_1=(0,0,1),\:\g{y}_2=(1,0,0),\nonumber\\
&&\g{y}_{3,4}=(-1/2,\pm\sqrt{3}/2,0)\nonumber
\ea
using the notation (\ref{sig}).
If the measured deviation squared is of the order of the variance then we cannot claim the higher state contribution.
For larger $W_k$ the formula (\ref{dev}) will reveal the estimated magnitude of the deviation.
As a side remark, this protocol applies also classically for $k=d$ to test contributions from higher classical states.

Generalization of the above protocol to higher $k$ is possible but more complicated. For instance, if $k=d^2+1$ or $d(d+1)/2$, the deviation is of higher order
\be
W_k=\sum_{ij,i'j'}\mathrm{sgn}(j'-j)\delta p_{ij}\delta p_{i'j'}M_{jj',ii'}(-1)^{i+j+i'+j'}
\ee
where $i<i'$, $j\neq j'$ and $M_{jj',ii'}$ is the minor obtained by removing columns $i,i'$ and rows $j,j'$.
On the other hand the variance in the null case (qubit) is
\be
N^2\langle W_k^2\rangle\simeq \sum_{ij,i'j'}p_{ij}p_{i'j'}(1-p_{ij})(1-p_{i'j'})M^2_{jj',ii'}\label{e11}
\ee
Equation (\ref{e11}) shows that detection nonzero $W_k$ of the given number of repetitions $N$ for $k=d^2+1$ or $d(d+1)/2+1$ has the same confidence level as for $d^2$ and $d(d+1)/2$ for small deviations $\delta p$.

As examples, for a qubit or qutrit we can take pure states $\hat{X}^{0}_j=|x_j\rangle\langle x_j|$ and $\hat{Y}^{0}_i=|y_i\rangle\langle y_i|$, and
the smallest deviations should have the forms $\delta\hat{X}_j=|x'_j\rangle\langle x_j|+|x_j\rangle\langle x'_j|$
and $\delta\hat{Y}_i=|y'_i\rangle\langle y_i|+|x_i\rangle\langle y'_i|$ with some small vectors $|x'_j\rangle$ and $|y'_i\rangle$
being either just purely  imaginary (for a real state test) or in higher level space (for a complex state test).
Let us consider a test of a qubit $\hat{X}^{0}_j=\hat{Y}^{0}_j$ defined by (\ref{sig}) with
\ba
&&\g{x}_1=-\g{x}_2=(1,0,0),\:\g{x}_3=(0,1,0)\nonumber\\
&&\g{x}_4=(0,0,1)=-\g{x}_5
\ea
which gives
\be
4W_4=\sum_{i=1,2}(\delta p_{i1}+\delta p_{i2}-\delta p_{i4}-\delta p_{i5})
\ee
and
$
\langle W_4^2\rangle\simeq 1/16N
$.
Certainly, our proposal does not exclude other choices, and the potential higher-level contributions can occur in different operations.
One should remember that the test can only falsify the assumed dimension since obtaining $W_k=0$ (within the error bounds) can happen in an accidental configuration.

\section{ Discussion}
An equality-based dimension witness with an optimal number states and measurements can  help in efficient diagnostics of the working Hilbert space of the supposed quantum systems. 
Deviations from zero can reveal the  influence of parasitic states. By testing various sets of preparations, one should quantify them and take measures to eliminate it. The witness can be generalized to various suboptimal 
configurations, e.g. combinations of more preparations and/or measurements. 
In any case, the applicability of this scheme can depend on the actual physical system and conjectured contributions from extra states. Even more generally, the witness can, in principle, test any
dimension-limited sub-algebra of the Hilbert space. At high dimensions, it becomes problematic to find maximal $W_k$, which is essential to estimate the sensitivity of the witness to extra-space contributions.

\section*{Acknowledgements}

J. B. acknowledges fruitful discussions with J. Rossell\'o, 
M. del Mar Batle and R. Batle. A.B. acknowledges discussion with J. Tworzyd{\l}o.
\appendix

\section{Numerical search for the maximum}

Our basic algorithm to find the maximum of $W_k$ (\ref{dett}) is as follows
\begin{footnotesize}
\begin{eqnarray}
1 &&\,\, initial\,set\,of\,states\,\{x_i\} \,and\,observables\,\{y_j\}\,given\cr
2 &&\,\, T \leftarrow T_0\cr
3 &&\,\, {\bf repeat \,until} \,stopping\, criterion \,is \,met\cr
4 &&\,\, ~~~~~~~{\bf repeat\,} (k,d)\,times \cr
5 &&\,\, ~~~~~~~~~~~~~~orientate \,all\, unit \,vectors \cr
6 &&\,\, ~~~~~~~~~~~~~~move \,to \,the \,next \,term \,within \,the \,set \cr
7 &&\,\, ~~~~~~~{\bf endrepeat}  \cr
8 &&\,\, ~~~~~~~set\,the\,matrix\,p \cr
9 &&\,\, ~~~~~~~compute \,W_k \cr
10&&\,\, ~~~~~~~T_{s+1} \,\leftarrow \,r\, T_s \cr
11&&\,\, {\bf endrepeat} \cr
12&&\,\, return \,the\,supremum\,of\,W_k\cr
\end{eqnarray}
\end{footnotesize}
\noindent The number of updates $(k,d)$ depends on the way the states are parametrized ($k$ the number of $x_i$ states, and $d$ is their dimension)

In order to illustrate the dynamics of the numerical analysis we shall reproduce the numerics of the case $k=4$ and a ququart (a four-level system, $d=4$) for two cases, namely, 
$\hat{Y}_j$  are (i) only one-dimensional projections $|y_j\rangle \langle y_j|$ and ii) as to two-dimensional projections $|y_j^{(1)}\rangle \langle y_j^{(1)}|+|y_j^{(2)}\rangle \langle y_j^{(2)}|$. 
In this last case, $|y_j^{(1)}\rangle$ and $|y_j^{(2)}\rangle$ ought to be orthogonal, so we must implement a Gram-Schmidt orthogonalization process. In both instances, 
we must maximize the absolute value of the determinant of the matrix $p_{ij}$
with $p_{ij}=\langle x_i|\hat{Y}_j|x_i\rangle$ for $i<5$ and $p_{5j}=1$

In principle, for $d=4$ the general parametrization for a complex ray $x_i$ requires $2\times4 = 8$ real numbers, which reduce to 6 after normalization and a 
global phase, which irrelevant for the computation of $p_{ij}$. However, this number can be lowered a bit further after some rotations (unitary operations).
Thus, we shall have

\begin{figure}[thbp!]
\begin{center}
\includegraphics[width=8.0cm]{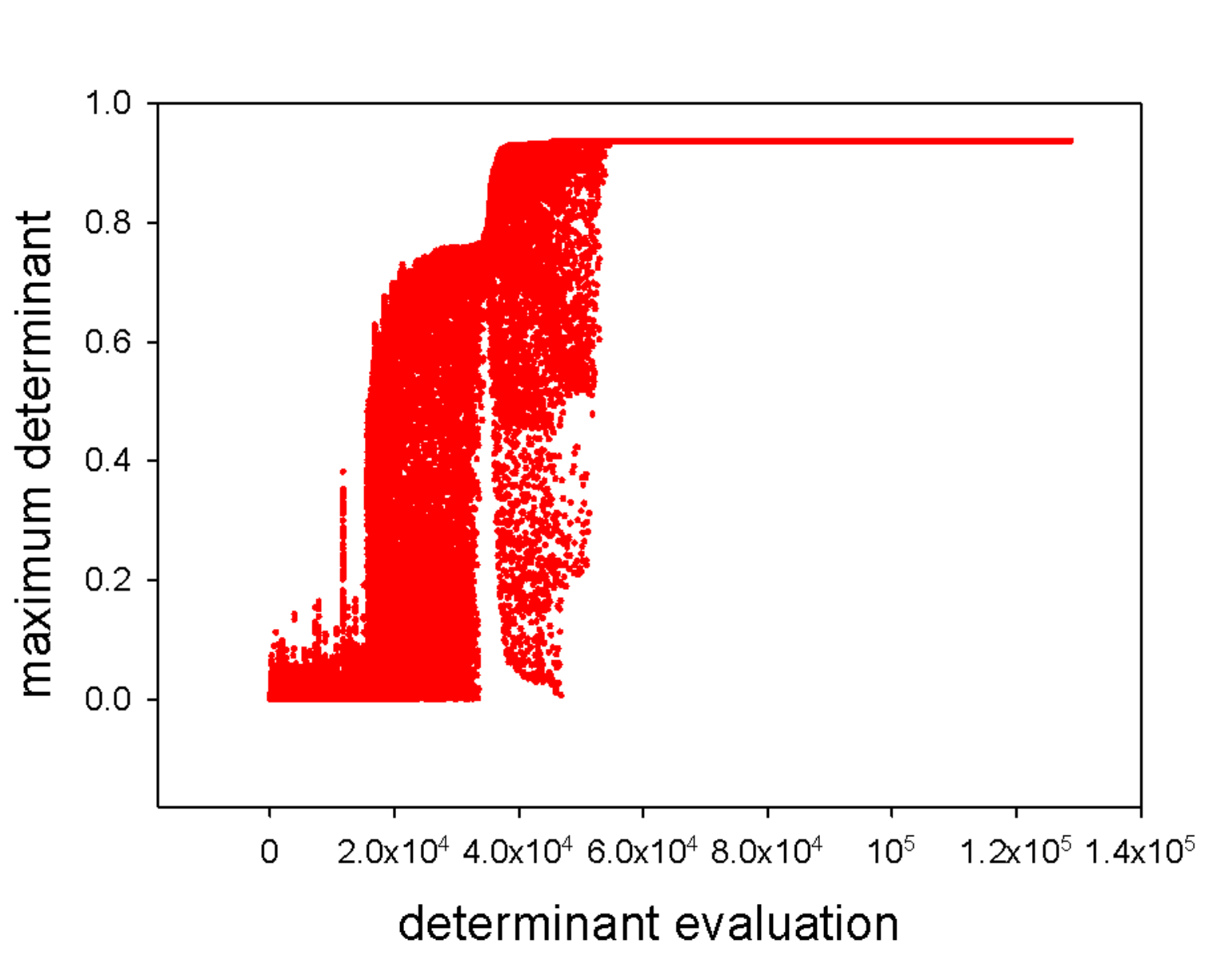}
\caption{Plot of the evolution of the maximum determinant for $k=5,d=4$ and one projector for the measurement, during the simulated annealing computation. See text for details.}
\label{figS1}
\end{center}
\end{figure}

\begin{figure}[thbp!]
\begin{center}
\includegraphics[width=8.0cm]{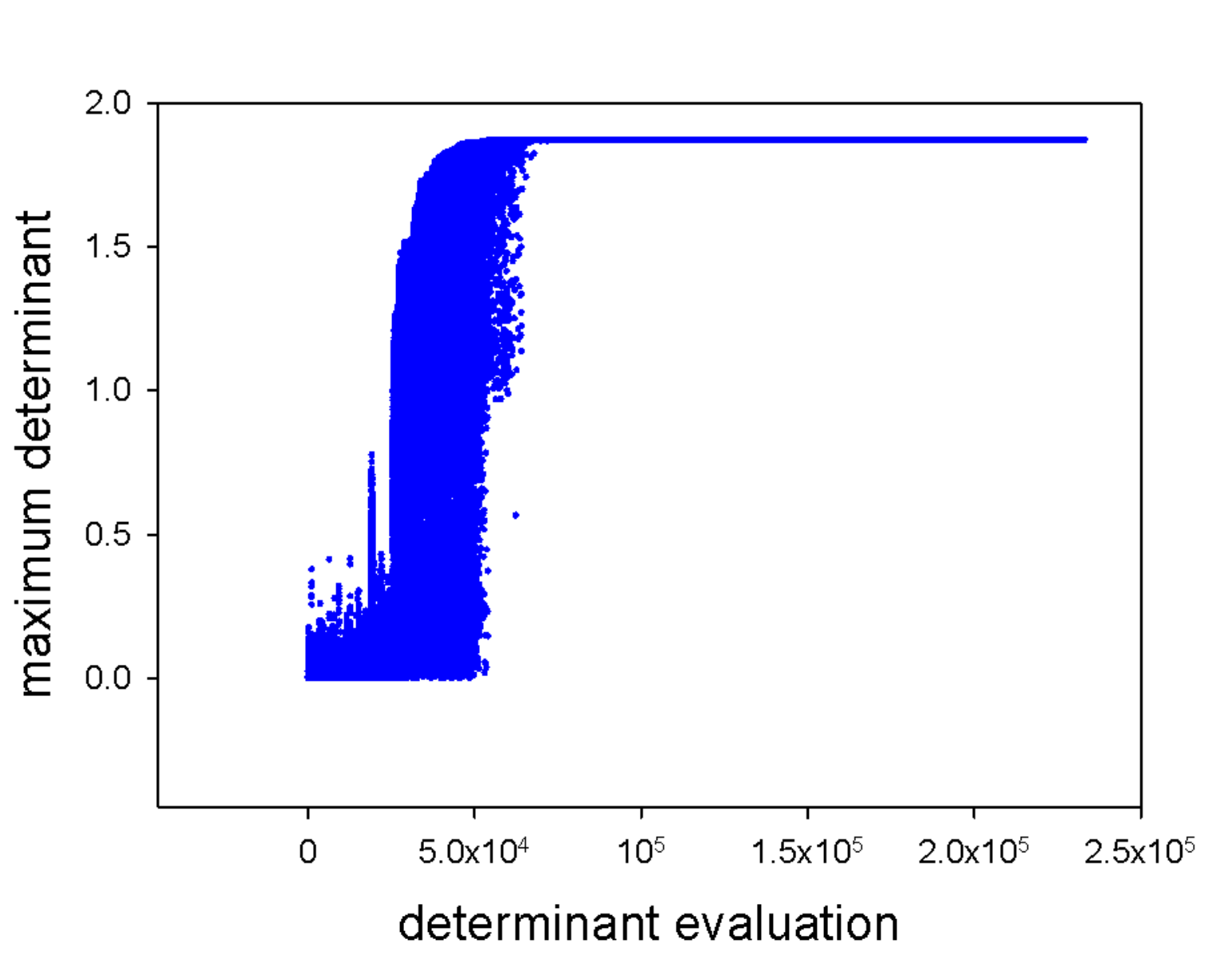}
\caption{Plot of the evolution of the maximum determinant for $k=5,d=4$ and two projectors for the measurement, during the simulated annealing computation. See text for details.}
\label{figS2}
\end{center}
\end{figure}

\begin{eqnarray} \label{xstates}
x_1&=&[1,0,0,0]\nonumber\\
x_2&=&[\cos A(1),\sin A(1),0,0]\nonumber\\
x_3&=&[\cos A(2) \cos A(3), \e^{i A(4)} \sin A(2)\cos A(3), \nonumber\\
      && \sin A(3), 0]\nonumber\\
x_4&=&[\sin A(5) \sin A(6) \cos A(7), 	\nonumber\\
      && \e^{i A(8)} \sin A(5) \sin A(6) \sin A(7),\\
      && \e^{i A(9)} \sin A(5) \cos A(6), \cos A(5)]\nonumber\\
x_5&=&[ \sin A(10) \sin A(11) \cos A(12),\nonumber\\
&& \e^{i A(13)} \sin A(10) \sin A(11) \sin A(12),\nonumber\\
&& \e^{i A(14)} \sin A(10) \cos A(11),\nonumber\\
&& \e^{i A(15)}\cos A(10)].\nonumber
\end{eqnarray}
\noindent The additional $\{y_1,y_2,y_3,y_4\}$ follow the general expression of state $x_5$ in (\ref{xstates}), adding a total of 39 angles, 
such that $A(j)\in [0,2\pi) $, for $j=1-39$. Similarly, for the case of two projectors, we have a total of 63 angles, that is, $A(j)\in [0,2\pi)$, for $j=1-63$.

In Fig.\ref{figS1} we depict the evolution of the value of $|W_k|$ versus each individual determinant evaluation over the set of 
$A(j)\in [0,2\pi)$, $j=1..39$ angles. 
The evolution of $|W_k|$ is bounded from above by the value 0.936442615 which, upon identification of the corresponding states $\{x_i,y_j\}$, easily leads to the analytic 
result $W_k = 2^{11}/3^7$. Likewise, the evolution of the value of $|W_k|$ for the case of two projectors is shown in Fig.\ref{figS2}, returning the maximum value 
$W_k = 2^{12}/3^7 \simeq 1.87288523 $.

Other instances are tackled exactly in the same vein, changing only the number of variables, which requires further computational effort.

\section{Special cases with the half-analytic representation}

For $k=2$ and $k=3$ and a qubit we can use the Bloch representation (4) writing
\be
p_{ij}=(1+\g{y}_i\cdot\g{x}_j)/2
\ee
so by a quick algebra we get
\ba
&& W_2=((\g{x}_1-\g{x}_3)\times(\g{x}_2-\g{x}_3))\cdot(\g{y}_1\times\g{y}_2)/4,\\
&&W_3=[\g{x}_1-\g{x}_4,\g{x}_2-\g{x}_4,\g{x}_3-\g{x}_4] [\g{y}_1,\g{y}_2,\g{y}_3]/8
\nonumber
\ea
where $[\g{a},\g{b},\g{c}]=\g{a}\cdot(\g{b}\times\g{c})$ is the mixed product.
Now with $|\g{x}_j|=|\g{y}_i|=1$ we quickly find that the maximum is achieved if
$\g{y}_i$ are the two/three axes while $|(\g{x}_1-\g{x}_3)\times(\g{x}_2-\g{x}_3)|/2$ is the area of the triangle and $[\g{x}_1-\g{x}_4,\g{x}_2-\g{x}_4,\g{x}_3-\g{x}_4]/6$ is the volume of the tetrahedron
with vertices  $\g{x}_j$. To get the largest volume, the triangle (tetrahedron) inscribed in the cricle (sphere) of radius $1$
must be regular. A simple argument is that the area (volume) can be always increased moving an apex to the diameter perpendicular to the opposite side (face),
making the edges adjacent to the apex equal.

For other cases we can guess the partial symmetry of the states and measurements configurations.

For $k=3$ and a qutrit, the optimal case (real and complex) is
\ba
&&|x_j\rangle=a\cos\frac{2\pi j}{3}|1\rangle+a\sin\frac{2\pi j}{3}|2\rangle)+b|3\rangle\nonumber\\
&&
|y_j\rangle=q\cos\frac{2\pi j}{3}|1\rangle+q\sin\frac{2\pi j}{3}|2\rangle+r|3\rangle
\ea
for $j=1,2,3$, and $|x_4\rangle=|3\rangle$ and with $a^2+b^2=q^2+r^2=1$ gives $
W_3 =(27/32)(4br+aq)^2(3q^2-2)a^4q^2$
The naive choice $a=q=1$ gives $27/32$ but the actual maximum $0.8447648009582842=1.0012\times 27/32$ is higher for
$a=0.993819$, $q=0.996329$.
 
For $k=4$ and a real qutrit, we take
\ba
&&|x_1\rangle=|1\rangle,\nonumber\\
&&|x_{2,3}\rangle=\frac{-|1\rangle\pm \sqrt{3}|2\rangle}{2},\:|x_{4,5}\rangle=\frac{-|1\rangle\pm\sqrt{3}|3\rangle}{2}\nonumber\\
&&
|y_{1,2}\rangle=a|1\rangle \pm b|2\rangle,\:|y_{3,4}\rangle=q|1\rangle\pm r|3\rangle
\ea
with $a^2+b^2=q^2+r^2=1$, $aq=1/2$
Then, $W_4=27\sqrt{5/4-z}(7/4-z)/8$
for $z=a^2+q^2$. From maximization we get $z=9/8$ and $W_4=27\sqrt{2}/64\simeq 0.5966213466261494$

When searching for the maximum for a complex qutrit, 
the numerical analysis shows that the maximal case lies in a particular subset of all possibilities  given by projection
 onto
 \ba
&&|x_j\rangle=a|1\rangle+b\omega^j|2\rangle+c\omega^{2j}|3\rangle,\:|x_4\rangle=|1\rangle,\:|x_5\rangle=|2\rangle\nonumber\\
&&|y_j\rangle= q|1\rangle+\omega^j r|2\rangle+\omega^{2j} s|3\rangle,\:|y_4\rangle=|1\rangle,
\ea
for $j=1,2,3$
with $a,b,c,p,q,r\in[-1,1]$, $a^2+b^2+c^2=q^2+r^2+s^2=1$.
Then
\be
\det p=27 c^2(s^2-r^2)( aqbr+brcs+csaq)^2
\ee
and maximizing over two spheres gives approximately $0.6319201017558774$.
For a ququart (a four-level system) the maximum is larger, and we can distinguish two cases, $\hat{Y}$ as
only one-dimensional projections and up to two-dimensional projections. In both cases we take
\be
\sqrt{3}|x_j\rangle=\sqrt{3}|y_1\rangle=\pm|1\rangle\pm|2\rangle\pm|3\rangle
\ee
for $j=1,2,3,4$, taking the cases with an even number of minus signs (again vertices of a regular tetrahedron) and $|x_5\rangle=|4\rangle$
Then for $\hat{Y}_j=|y_j\rangle\langle y_j|$ we get $\det=2^{11}/3^7\simeq 0.94$ while for
$\hat{Y}_j=|y_j\rangle\langle y_j|+|4\rangle\langle 4|$
we get $W_4=2^{12}/3^7\simeq 1.872885230909922$.

For $k=5$, we can maximize $\det$ with a real qutrit, with $x$ and $y$ being in the independent bases, with
\be
|x_{3a+b}\rangle=(|a\rangle+(-1)^b\phi|a+1\rangle)/\sqrt{\phi+2}
\ee
for $a=0,1,2$ and $0\equiv 3$, $b=1,2$, and the golden ratio $\phi=(1+\sqrt{5})/2$, i.e., (pairs of) vertices of a regular icosahedron,
while

\be
|y_j\rangle=\alpha(\cos(2\pi j/5)|1\rangle+\sin(2\pi j/5)|2\rangle)+\sqrt{1-\alpha^2}|3\rangle.
\ee
We have to maximize $(t-1)(3t-2)t^3$, $\alpha=t^2$,
which has the derivative
$
t^2(15t^2-20t+6)
$
with the nontrivial roots $t=(10\pm\sqrt{10})/15$. The larger $W_5=(25 + 34 \sqrt{10}) 2^5/5^3 3^4\simeq 0.4188205525198219$ is for $+$.

A complex qutrit gives a larger maximum $0.457413503$ in the real $10$-parameter family of vectors
\ba
&&|y_1\rangle=|1\rangle,\:|x_{j}\rangle=a_{j}|1\rangle+b_{j}|2\rangle\\
&&|x_{34}\rangle=f|1\rangle+g|2\rangle\pm h|3\rangle,\nonumber\\
&&|x_{56}\rangle=f'|1\rangle+g'|2\rangle\pm ih'|3\rangle\nonumber\\
&&|y_{23}\rangle=q|1\rangle+r|2\rangle\pm s|3\rangle,\nonumber\\
&&|y_{45}\rangle=q'|1\rangle+r'|2\rangle\pm is'|3\rangle\nonumber
\ea
for $j=1,2$,
with $a_j^2+b_j^2=f^2+g^2+h^2=q^2+r^2+s^2=f^{\prime 2}+g^{\prime 2}+h^{\prime 2}=q^{\prime 2}+r^{\prime 2}+s^{\prime 2}$
(two circles and four spheres).

For a ququart, the maximum is found for two-dimensional $\hat{Y}_j$ (real and complex).
We take in the 
\ba
&&|x_{12}\rangle=a|1\rangle\pm b|3\rangle,\nonumber\\
&&|x_{34}\rangle=a|3\rangle\pm b|4\rangle),\nonumber\\
&&|x_{56}\rangle=(|2\rangle\pm |4\rangle)/\sqrt{2}
\ea

and $\hat{Y}_j=|y_j\rangle\langle y_j|+|y'_j\rangle\langle y'_j|$,
\ba
&&|y_1\rangle=|y_2\rangle=(|1\rangle+|2\rangle)/\sqrt{2},\:|y_3\rangle=|1\rangle,\nonumber\\
&&|y_{45}\rangle=s|2\rangle \pm c|4\rangle,\nonumber\\
&&|y'_{12}\rangle=(|3\rangle\pm|4\rangle)/\sqrt{2},\nonumber\\
&&|y'_{345}\rangle=|3\rangle,
\ea
with $a^2=(5+\sqrt{5})/10$, $b^2=(5-\sqrt{5})/10$ and $c=\cos(\pi/5)=(1+\sqrt{5})/4$, $s=\sin(\pi/5)=\sqrt{(5-\sqrt{5})/8}$.
Then
\be
W_5=(1+1/\sqrt{5})^{5/2}/\sqrt{2}\simeq 1.78162618305857
\ee

For a real and complex ququint ($d=5$)
and the case of single projections $\hat{Y}_j=|y_j\rangle\langle y_j|$, we take
 the vertices of a 5-cell (a generalization of a tetrahedron in four dimensions)  in space $1234$
\ba
&&|x_1\rangle=|4\rangle,\:|x_6\rangle=|5\rangle\nonumber\\
&&|x_2\rangle=[\sqrt{5}(|1\rangle+|2\rangle+|3\rangle)-|4\rangle]/4\nonumber\\
&&|x_3\rangle=[\sqrt{5}(|1\rangle-|2\rangle-|3\rangle)-|4\rangle]/4,\nonumber\\
&&
|x_4\rangle=[\sqrt{5}(|2\rangle-|3\rangle-|1\rangle)-|4\rangle]/4\nonumber\\
&&
|x_5\rangle=[\sqrt{5}(|3\rangle-|1\rangle-|2\rangle)-|4\rangle]/4,
\ea
and $|y_j\rangle=|x_j\rangle$,
giving $W_5=5^5 3^4/2^{18}$.

The case of double projections $\hat{Y}_j=|y_j\rangle\langle y_j|+|y'_j\rangle\langle y'_j|$
for
\ba
&&|x_j\rangle=a|1\rangle+\cos\frac{2\pi j}{3}b|2\rangle+\sin\frac{2\pi j}{3}b|3\rangle,\nonumber\\
&&
|y_j\rangle=q|1\rangle-\cos\frac{2\pi}{3}r|2\rangle-\sin\frac{2\pi j}{3}r|3\rangle,\nonumber\\
&&|y'_j\rangle=\sin\frac{2\pi j}{3}|2\rangle-\cos\frac{2\pi j}{3}|3\rangle,\nonumber\\
&&|x_4\rangle=|y_4\rangle=|4\rangle,\:|x_5\rangle=|y_5\rangle=|5\rangle,\\
&&|x_6\rangle=|y'_4\rangle=|y'_5\rangle=|1\rangle,\nonumber
\ea
for $j=1,2,3$, with $a^2+b^2=q^2+r^2=1$.
Then $W_5=(27/32) b^4q^2(2+q^2)(4ar+bq)^2$ .
Maximizing with respect to $b$ at constant $c$ and $a$ gives the condition
\be
c^2=\frac{(8-12b^2)^2}{(8-12b^2)^2+9b^2(1-b^2)}
\ee
Inserting it into the formula we get
\be
W_5=\frac{6^4 b^6 ( 3 b^2-2)^2 (32 - 93 b^2 + 69 b^4)}{(64 - 183 b^2 + 135 b^4)^3}
\ee
whose maximum $3.144615108566082$ is for the largest root of 
\be
2048 - 12032 b^2 + 26409 b^4 - 25668 b^6 + 9315 b^8=0
\ee

For $k=6,7$ and $d=3$ the complex case gives the respective maxima $0.330364646,0.225213334$.
For $k=6$ and $d=4$ in the real and complex cases we got $1.61439616$ and $1.68093981$, respectively, with two-dimensional $\hat{Y}$.
For $d=5$ in the real case we get  $3.39847186$, and the complex case is the same.
In this case the maximum is realized for
\ba
&&|x_{12}\rangle=a|1\rangle\pm b|2\rangle,\nonumber\\
&&|x_{34}\rangle=a|1\rangle\pm b|3\rangle,\nonumber\\
&&|x_{56}\rangle=a|1\rangle\pm b|4\rangle,\nonumber\\
&&|x_7\rangle=|5\rangle,
\ea
with $a^2+b^2=1$, while $\hat{Y}_i=|y_i\rangle\langle y_i|+|y'_i\rangle\langle y'_i|$
with
\ba
&&|y_{12}\rangle=c|1\rangle\pm d|1\rangle,|y'_{12}\rangle=|3\rangle,\nonumber\\
&&|y_{34}\rangle=c|1\rangle\pm d|3\rangle,|y'_{34}\rangle=|4\rangle,\\
&&|y_{56}\rangle=c|1\rangle\pm d|4\rangle,|y'_{56}\rangle=|2\rangle,\nonumber
\ea
with $c^2+d^2=1$ giving $W_6=512 a^3b^7c^3d^3(3a^2c^2+(1+d^2)b^2)(c^2+d^4)$
The maximum is $3.3984718576415207$ for $a$ being the root of the equation
\ba
&&0=-1 + 5 a^2 + 24 a^4 - 296 a^6+ 1480 a^8\\
&& - 5088 a^{10} + 11392 a^{12} - 
 14336 a^{14} + 8192 a^{16},\nonumber
\ea
$a^2=0.3016492773799042$.
For $d=6$ we get the same maximum in the real and complex case for states
\ba
&&|x_1\rangle=a|1\rangle+b|2\rangle,\:|x_{23}\rangle=a|1\rangle+\frac{b}{2}(-|2\rangle\pm\sqrt{3}|2\rangle),\nonumber\\
&&|x_4\rangle=a|1\rangle+b|4\rangle,\:|x_{56}\rangle=a|1\rangle+\frac{b}{2}(-|4\rangle\pm\sqrt{3}|5\rangle),\nonumber\\
&&|x_7\rangle=|6\rangle
\ea
while $\hat{Y}_j=|y_j\rangle\langle y_j|+|y'_j\rangle\langle y'_j|+|6\rangle\langle 6|$ with
\ba
&&|y_1\rangle=|3\rangle,\:|y_{23}=|3\rangle/2\pm\sqrt{3}|2\rangle/2,\\
&&|y_4\rangle=|5\rangle,\:|y_{23}=|5\rangle/2\pm\sqrt{3}|4\rangle/2,\nonumber\\
&&|y'_1\rangle=c|1\rangle+d|2\rangle,\:|y'_{23}\rangle=c|1\rangle+\frac{d}{2}(-|2\rangle\pm\sqrt{3}|2\rangle),\nonumber\\
&&|y'_4\rangle=c|1\rangle+d|4\rangle,\:|y'_{56}\rangle=c|1\rangle+\frac{d}{2}(-|4\rangle\pm\sqrt{3}|5\rangle),\nonumber
\ea
for $a^2+b^2=c^2+d^2=1$ giving
\ba
&&W_6=\frac{3^6}{2^{10}}b^6c^2(1+d^2)(2+c^2+2a^2-5a^2c^2)\times\nonumber\\
&&(c^4+98a^2d^2(c^2-a^2)-2a^2c^2+a^4+343a^4d^2\nonumber\\
&&-16abc^3d-16a^3bcd(17d^2-1))
\ea
giving the numerical maximum $5.0467662420644475$.

For $k=7$ and $d=4$  in the real and complex cases, we get $W_7=1.41149223$ and $1.63898287$, respectively (two-dimensional $\hat{Y}$), respectively.
For $d=5$ in the real  and complex cases (two-dimensional $\hat{Y}$) we get $3.50557203$ and  $7^7/2^{13}3^3=3.723338939525463$, respectively.
In the latter case the states read $|x_8\rangle=|5\rangle$, while the rest are, in the basis $|1\rangle,|2\rangle,|3\rangle,|4\rangle$,
\ba
&&
|x_j\rangle=\frac{\hat{U}^j}{2}\begin{pmatrix}
1\\
1\\
1\\
1\end{pmatrix},\:\hat{U}=\begin{pmatrix}
1&0&0&0\\
0&\zeta&0&0\\
0&0&\zeta^2&0\\
0&0&0&\zeta^4\end{pmatrix},\nonumber
\\
&&
\hat{Y}_j=\hat{U}^j\begin{pmatrix}
1/2&\xi&\xi&\xi\\
-\xi&1/2&\xi&-\xi\\
-\xi&-\xi&1/2&\xi\\
-\xi&\xi&-\xi&1/2
\end{pmatrix}\hat{U}^{\dag j}
\ea
for $\zeta=\exp(2\pi i/7)$ and $\xi=i/2\sqrt{3}$.
Then, due to the Gauss identity
\be
1+2(\zeta+\zeta^2+\zeta^4)=i\sqrt{7}
\ee
we obtain
\be
|\langle x_i|x_j\rangle|^2=\left\{\begin{array}{ll}
1&\mbox{ for }i=j\\
1/8&\mbox{ for }i\neq j\mbox{ and }i,j<8\\
0&\mbox{ otherwise, }
\end{array}\right.
\ee
$\mathrm{Tr}\,\hat{Y}_j\hat{Y}_i=5/6$ for $i\neq j$,
and
\ba
&&p_{ji}\langle x_i|\hat{Y}_j|x_i\rangle=\\
&&=\left\{\begin{array}{ll}
1&\mbox{ for }j=8\\
0&\mbox{ for }i=8\mbox { and }j<8\\
1/2&\mbox{ for }i=j<8\\
\frac{1}{2}+\frac{\sqrt{7}}{4\sqrt{3}}&\mbox{ for } i,j<8\mbox{ and }j-i=1,2,4\mbox{ mod }7\\
\frac{1}{2}-\frac{\sqrt{7}}{4\sqrt{3}}&\mbox{ otherwise }
\end{array}\right.\nonumber
\ea
For $d=6$ in the real and complex cases, we get $6.05145518$ and $6.17876168$. 

We can maximize $W_8$ with a complex qutrit with $x$ and $y$ with optimal $x$:
\be
|x_{3a+b}\rangle=(|a\rangle +e^{i\alpha_a}\omega^b|a+1\rangle)/\sqrt{2}
\ee
for $\omega=e^{2\pi i/3}=(i\sqrt{3}-1)/2$, $a=0,1,2$, $b=1,2,3$ and $|0\rangle\equiv |3\rangle$ with free angles $\alpha_a$.
and optimal $y$:
\be
|y_j\rangle=\sqrt{5/6}|y'_j\rangle+\sqrt{1/6}|3\rangle
\ee
with $y'_j$ defined in $|1\rangle$, $|2\rangle$ space
\be
|y'_j\rangle=i\sqrt{1/3}|1\rangle+\omega^j\sqrt{2/3}|2\rangle
\ee
for $j=1,2,3$, $|y'_4\rangle=|1\rangle$ and $|y'_{j+4}\rangle=-|y'_j\rangle$
Then  the maximum is $W_8=5^5/3^4 2^8\simeq 0.1507040895061728$.

For $d=4$ in the real and complex cases,  $W_8=1.2962761$ and $1.47025989$,  while
for $d=5$ in the real and complex cases $W_8=3.64938453$ and $3.79495481$, and for $d=6$
in the real and complex cases $7.48985655$ and $7.49786979$, respectively.

The next nontrivial case is $k=9$ which discriminates between a qutrit and ququart. For instance, taking $\hat{Y}_{j-1}=\hat{X}_j$
with
\ba
&&|x_j\rangle=|j\rangle,\: j=1..4\\
&&2|x_j\rangle=\sum_{m=1}^4 i^{mj}|m\rangle,\:j=5..8\nonumber\\
&&2|x_9\rangle=|1\rangle+|2\rangle-|3\rangle-4\rangle\nonumber\\
&&2|x_{10}\rangle=|1\rangle-|2\rangle-|3\rangle+|4\rangle\nonumber
\ea
we get $W_9=1/8$.
However, the numerical maximum for $d=4$ is $W_9=1.28868526$ and $1.39037781$ for the real and complex case, respectively.
For $d=5$ we get respectively $3.76568067$ and $3.83579182$ and for $d=6$, we get $10.1361814$ and $10.3359304$.


\begin{thebibliography}{99}

\bibitem{gallego}
R. Gallego, N. Brunner, C. Hadley, and A. Acin, 
{\it Device-Independent Tests of Classical and Quantum Dimensions},
Phys. Rev. Lett. 105, 230501 (2010)

\bibitem{hendr}
M. Hendrych, R. Gallego, M. Micuda, N. Brunner, A. Acin, J. P. Torres,
{\it Experimental estimation of the dimension of classical and quantum systems},
Nat. Phys. 8, 588 (2012)

\bibitem{ahr}
J. Ahrens, P. Badziag, A. Cabello, and M. Bourennane, {\it Experimental Device-independent Tests of Classical and Quantum Dimensionality}
Nature Phys. 8, 592 (2012).

\bibitem{ahr2}
J. Ahrens, P. Badziag, M.Pawlowski, M. Zukowski, M. Bourennane,
{\it Experimental Tests of Classical and Quantum Dimensions},
Phys. Rev. Lett. 112, 140401 (2014)


\bibitem{dim1}
N. Brunner, M. Navascues, and T. Vertesi,
{\it Dimension Witnesses and Quantum State Discrimination}
Phys. Rev. Lett. 110, 150501 (2013)


\bibitem{leak}
A. Strikis, A. Datta, G. C. Knee,
{\it Quantum leakage detection using a model-independent dimension witness},
Phys. Rev. A 99, 032328 (2019)

\bibitem{dim}
J. Bowles, M. T. Quintino, and N. Brunner,
{\it Certifying the Dimension of Classical and Quantum Systems in a Prepare-and-Measure
Scenario with Independent Devices}, Phys. Rev. Lett. {\bf 112}, 140407 (2014)

\bibitem{chen}
X. Chen K. Redeker, R. Garthoff, W. Rosenfeld, J. Wrachtrup, and I. Gerhardt, {\it Certified randomness from remote state preparation dimension witness},
Phys. Rev. A 103, 042211 (2021)

\bibitem{real}
B. G. Christensen, Y.-Ch. Liang, N. Brunner, N. Gisin, and P. G. Kwiat
{\it Exploring the Limits of Quantum Nonlocality with Entangled Photons}
Phys. Rev. X 5, 041052 (2015)


\bibitem{sorkin}
R. Sorkin,{\it Quantum mechanics as quantum measure theory}, Mod. Phys. Lett. A \textbf{9}, 3119 (1994).

\bibitem{tslit}
U. Sinha, C. Couteau, T. Jennewein, R. Laflamme,and G. Weihs, 
{\it Ruling out multi-order interference in quantum mechanics},
Science  {\bf 329}, 418 (2010);


\bibitem{btest1}
D. K. Park, O. Moussa, and R. Laflamme, {\it
Three path interference using nuclear magnetic resonance: A test of the consistency of Born's rule},
New J. Phys. \textbf{14},  113025 (2012).

\bibitem{btest2}
M.-O. Pleinert, J. von Zanthier, and E. Lutz,
{\it Many-particle interference to test Born's rule},
Phys. Rev. Research 2, 012051(R) (2020)

\bibitem{born}
M. Born, {\it Quantenmechanik der Stossvorg{\"a}nge},
 Z. f. Physik 37, 863 (1926)


\bibitem{ehr}
H. Ehlich,{ \it Determinantenabsch{\"a}tzungen f{\"u}r bin{\"a}re Matrizen}, Math. Z., 83, 123 (1964),
\bibitem{woj}
M. Wojtas, {\it On Hadamard's inequality for the determinants of order non-divisible by 4}, Colloq. Math., 12, 73
(1964)
\bibitem{sloane}
N.J.A. Sloane, The on-line encyclopedia of integer sequences, A003432, https://oeis.org/A003432


\bibitem{kirkpatrick83} S. Kirkpatrick, C. D. Gelatt Jr., and M. P. Vecchi, Science {\bf 220}, 671 (1983).

\bibitem{Avriel} M. Avriel, {\it Nonlinear Programming: Analysis and Methods} (Dover, Mineola, 2003). 





\end{thebibliography}
\end{document}